\documentclass[12pt,twoside]{article}
\usepackage{fleqn,espcrc1,rotating}
\usepackage{latexsym} 
\usepackage{amssymb}  
\usepackage{amsfonts} 
\usepackage{multirow}
\usepackage{epsf}
\usepackage{graphicx}
\def\y0{y^{(0)}}

\newcommand \beq{\begin{eqnarray}}
\newcommand \eeq{\end{eqnarray}}

\newcommand{\mnote}[1]{\marginpar{\tiny {}}}   

\begin{document}

\title{\bf On Charm Production near the Phase Boundary
}
\author{
   Peter Braun-Munzinger\\
   Gesellschaft f{\"u}r Schwerionenforschung\\
   64220 Darmstadt, Germany\\
   Johanna Stachel\\
   Physikalisches Institut, Universit\"at Heidelberg\\
   69120 Heidelberg,Germany
   }
\maketitle

\begin{abstract}

    \noindent We discuss aspects of the statistical hadronization
model for the production of mesons with open and hidden charm in
ultra-relativistic nuclear collisions. Emphasis is placed on what can
be inferred from the  dependence of the yield of charmonia on the
number of participants in the collisions.

\end{abstract}


\section{Introduction} 
\noindent Experiments  with ultra-relativistic nuclei are performed to
produce and study the quark-gluon plasma. This new state of
matter is predicted to exist at high temperatures and/or high baryon
densities. Numerical solutions of QCD using lattice
techniques imply that the critical temperature (at zero baryon
density) is about 170 MeV \cite{lattice}. Comprehensive surveys of the
various experimental approaches on how to produce such matter in
nucleus-nucleus collisions have been given recently
\cite{pbmpanic,jsinpc,pbmqm97,bass}.  Here we focus  on
charm  production and  its recent  interpretation in terms of a
statistical hadronization model \cite{jpsi-stat}. Since the
statistical hadronization model for charm is a consequence of the
success of thermal model descriptions for hadrons produced in nuclear
collisions we first very briefly review the present state there. We
then present an update of the extension of this model to charm
production with emphasis on  what can
be inferred from the dependence of the yield of charmonia on the
number of participants in the collisions. Remarks on what can be
expected at collider energies will conclude the manuscript.

\section{Equilibration at the Phase Boundary}
\begin{figure}[thb]

\vspace{-1cm}

\epsfxsize=15cm
\begin{center}
\hspace*{0in}
\epsffile{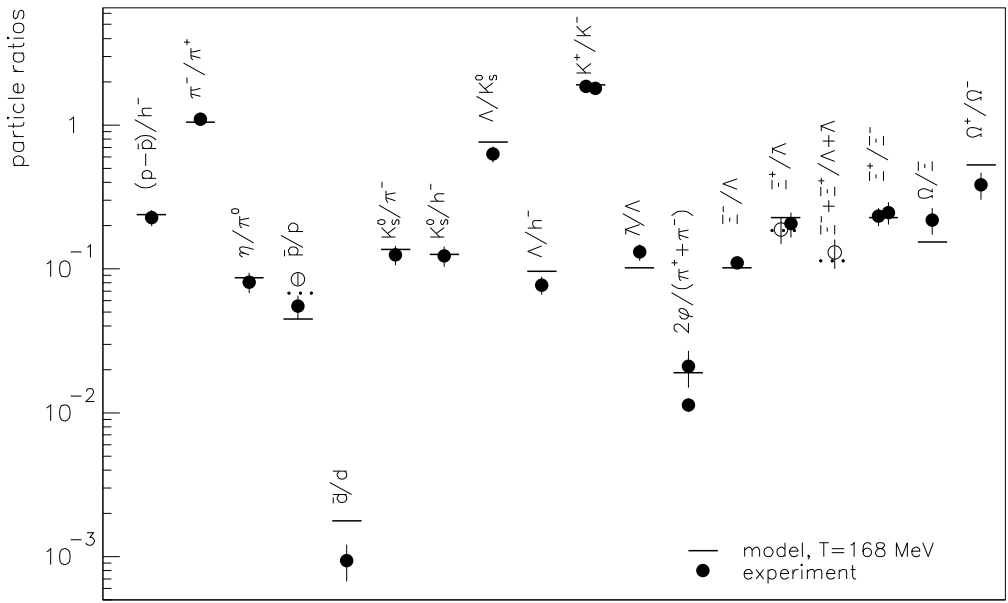}
\end{center}

\vspace{-1.5cm}

\caption{
Comparison of measured particle ratios with predictions of the thermal
model. For details see text and \cite{therm3}.
} 
\label{fig:ratios}
\end{figure}

\noindent 
The statistical model used to describe hadron multiplicities is
presented in detail in 
\cite{therm3}. Like its predecessors  presented
in \cite{therm2,therm1} it is based on the use of a grand canonical ensemble
to describe the partition function and hence the density of the
hadrons under consideration.  Without invoking volume information this
model can then be used to describe ratios of particle yields.
The surprizing result of applying this very simple model to data is
that ratios of all hadron yields including those involving
multi-strange baryons, where enhancement factors of more than 1 order
of magnitude over what had been measured in p+Be collisions are
observed,  can be described consistently 
\cite{therm3} if one assumes a 
fireball with temperature T=168 MeV and baryon chemical potential
$\mu_B$ = 266 MeV. This is demonstrated in Fig.~\ref{fig:ratios}. On
the other hand, the observed enhancement, especially for 
multistrange hadrons, cannot currently  be understood within any of the
hadronic event generators \cite{qm99}

\begin{figure}[thb]

\vspace{-3.1cm}

\epsfxsize=10cm
\begin{center}
\hspace*{0in}
\epsffile{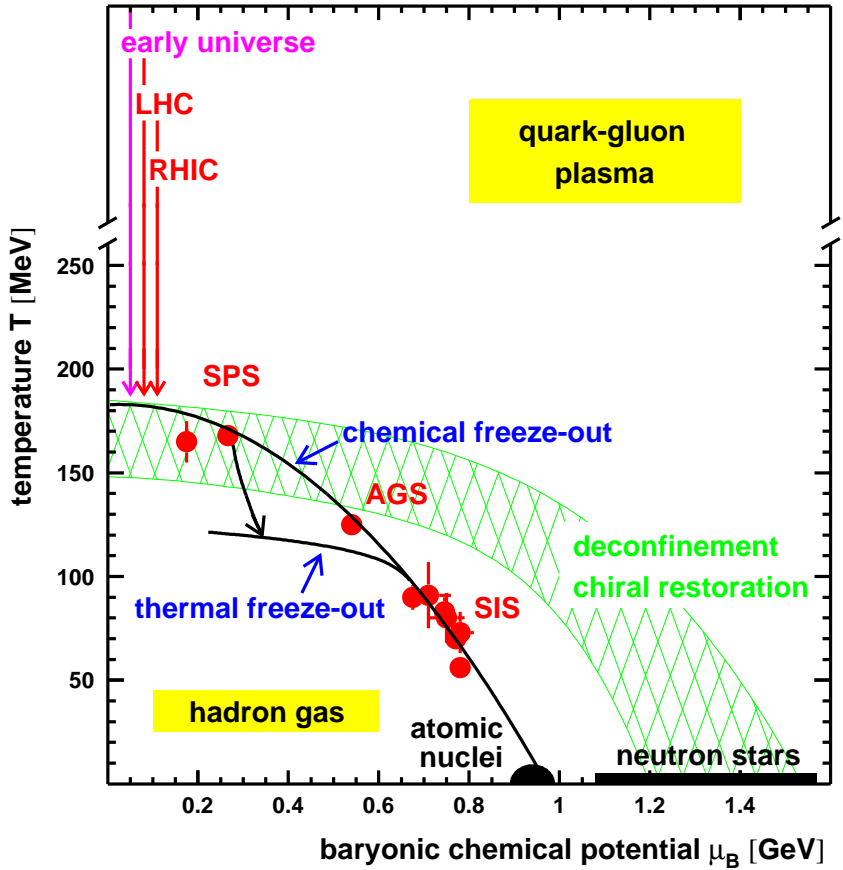}
\end{center}

\vspace{-2.0cm}

\caption{
Phase diagram of hadronic and partonic matter. The hadrochemical
freeze-out points are determined from thermal model analyses of heavy
ion collision data at SIS, AGS and SPS energy. The hatched region
indicates the current expectation for the phase boundary based on lattice QCD
calculations at $\mu_B$=0. The arrow from chemical to thermal
freeze-out for the SPS corresponds to isentropic expansion.
} 
\label{fig:phase-d}
\end{figure}

The chemical
potentials $\mu_B$ and temperatures T resulting from the such thermal
analyses  
\cite{therm3,therm2,therm1} place the systems at chemical freeze-out
very close to where we currently believe is the phase boundary between
plasma and hadrons. This is demonstrated in Fig.~\ref{fig:phase-d}.
The freeze-out trajectory (solid curve through the data
points) is  to guide the eye.

The closeness of the freeze-out parameters (T,$\mu_B$) to the phase
boundary might be the clue to the apparent chemical equilibration in
the hadronic phase: if the system prior to reaching freeze-out was in
the partonic (plasma) phase, then hadron production in general and
strangeness production in particular is determined 
by larger partonic cross sections as well as by hadronization.

\section{Statistical Hadronization of Charm}
\noindent Charm quarks are heavy ($m_c \gg T_c$) and thermal
production of charm quarks and charmed hadrons is not likely in
ultra-relativistic nuclear collisions. The situation has been recently
discussed \cite{jpsi-stat} with the conclusion that, compared to direct
hard production, thermal production of charm quarks can be neglected at
SPS energies and is small even at LHC energy. However, in the course
of these investigations we proposed a new scenario for quarkonia and
charmed hadron production. We assume \cite{jpsi-stat} that {\bf
all} c$\bar {\rm c}$ pairs are produced in direct, hard collisions,
i.e. in line with the above considerations we neglect thermal
production. For a description of the hadronization of the c and $\bar
{\rm c}$ quarks, i.e. for the determination of the relative yields of
charmonia, and charmed mesons and baryons, we employ the statistical
model, with parameters as determined by the analysis of all other
hadron yields \cite{therm3}. The picture we have in mind is that all
hadrons form within a narrow time range at or close to the phase
boundary. 

Since the number of directly produced charm quarks deviates from the
value determined by chemical equilibration, we introduced a charm
enhancement factor $g_c$ by the requirement of charm
conservation. This leads to the following relation between hard open
charm production and thermal production:

\beq
{N_{c\bar c}^{direct} = \frac{1}{2} g_c(\sum_{i}
N_{D_i}^{therm}+N_{\Lambda_i}^{therm})+
g_c^2(\sum_{i}N_{ \psi_i}^{therm}) +...}.
\eeq

Consequently, the charm enhancement factor $g_c$ can be determined if
$N_{c\bar c}^{direct}$ is known. Since there are no data yet for open
charm production in nuclear collisions we rely, as discussed in
\cite{jpsi-stat}, on the investigations and model calculations of
\cite{misko}, where the 
information available on charm production in hadron-nucleus collisions
is extrapolated to Pb-Pb collisions at SPS energy.
The number of  J/$\psi$ mesons is then enhanced relative to the
thermal model prediction by a factors of  g$_c^2$, i.e. 

\beq
N_{J/\psi}=g_c^2 N_{J/\psi}^{therm}.
\eeq

\begin{figure}[thb]

\vspace{2.5cm}

\epsfxsize=11cm
\begin{center}
\begin{turn}{270}
\hspace*{-2.0cm}
\epsffile{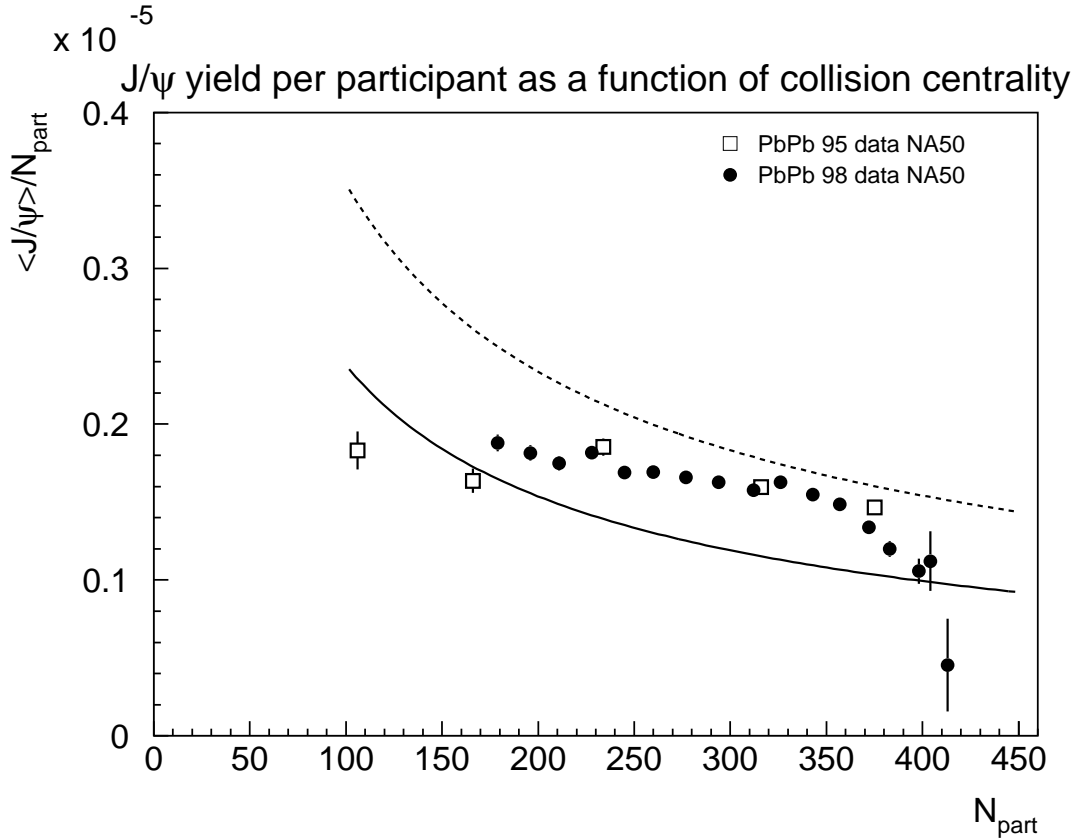}
\end{turn}
\end{center}

\vspace{-1.5cm}

\caption{ Comparison of the dependence of the measured
(J/$\psi)$/N$_{part}$ ratio 
on the number of participating nucleons with the predictions of the
of the canonical direct/statistical model (solid
line). The data are from a reanalysis of 
\cite{na50_1,na50_2,cicalo} and are normalized using the
(J/$\psi$)/h$^-$ ratio deduced in \cite{gaz2}. For
details, in particular about the normalization,  see text. The dashed
line is obtained by increasing the direct open charm yield by 50~\%
}
\label{fig:psipp}

\end{figure}

Application of this model to the data for charmonium production is
somewhat complicated since the NA50 collaboration has not provided
J/$\psi$ yields as a function of centrality or number of participants
in Pb+Pb collisions at SPS energy. Gosset et al. \cite{gosset1} have
provided such an analysis based on preliminary 1995 data and this was
the basis of 
the discussion in ref. \cite{jpsi-stat}. We will, for the following
discussion, mainly focus on the dependence of J/$\psi$ production on
the number of participating nucleons. The shape of this dependence can
be analyzed from published NA50 data \cite{na50_1,na50_2,cicalo}. The
absolute normalization we obtained by using for N$_{part} = 100$ that
N$_{J/\psi}/N_{h-}$ = 1.1$\cdot 10^{-6}$ \cite{gaz2}, leading to
N$_{J/\psi}$/N$_{part}$ = 1.83 $\cdot 10^{-6}$. These results agree
well with a recent analysis \cite{gosset2} of NA50 data of 1996 and
1998 within the framework of ref. \cite{gosset1} and are displayed in 
Fig.~\ref{fig:psipp}. We note that the so obtained data agree with the
analysis reported in \cite{gosset1} for low centralities but exceed
those data by about 30~\% for N$_{part}$=350.

In the following we will discuss these data within the framework of
ref. \cite{jpsi-stat}, with the following modifications. We take into
account, in the thermal model description, the full set of charmed mesons and
baryons. This increases, as mentioned already in \cite{jpsi-stat},
the (grand-canonical) thermal charm yield by 
about a factor 2.5, mostly because of the large statistical factors of the
D$^*$ mesons. Since, despite this increase, the number of charm pairs
is significantly less than 1 per collision, we treat the system, as
proposed by Redlich \cite{redlich}, within canonical
thermodynamics, following \cite{cleymans_redlich}. A similar approach
was recently chosen by \cite{gorenstein}, to investigate limits on
open charm production which are imposed within the present model by
the data on J/$\psi$ production.

Neglecting the (very small) quadratic terms in eq. (1) the statistical
hadronization model then reads:

\beq
{N_{c\bar c}^{direct} = \frac{1}{2} g_c 
N_{oc}^{therm} \frac{I_1(g_c N_{oc}^{therm})}{I_0(g_c N_{oc}^{therm})}}.
\eeq

Here, $N_{oc}^{therm}$ denotes the total thermal yield per collision of open
charm mesons and baryons (calculated within the grand-canonical
ensemble) and $I_1$ and $I_0$ are modified Bessel functions. From the
properties of the Bessel functions one obtaines $\lim_{x \rightarrow
\infty} \frac{I_1(x)}{I_0(x)}=1$ so  that the
grand-canonical limit is obtained for $g_c N_{oc}^{therm} \gg 1$.

Since quarkonia are hidden charm mesons, eq. (2) is still valid. To
set the stage for the following discussion we now proceed to evaluate
the dependence of J/$\psi$ production on N$_{part}$. For all estimates
we assume $N_{c\bar c}^{direct} \propto N_{part}^{(4/3)}$. Since
N$_{oc}^{therm} \propto N_{part}$ we get 
that:

\beq
{\frac{N_{J/\psi}}{N_{part}} \propto g_c^2}.
\eeq

\begin{figure}[thb]

\vspace{-8.5cm}

\epsfxsize=18cm
\begin{center}
\hspace*{1.0cm}
\epsffile{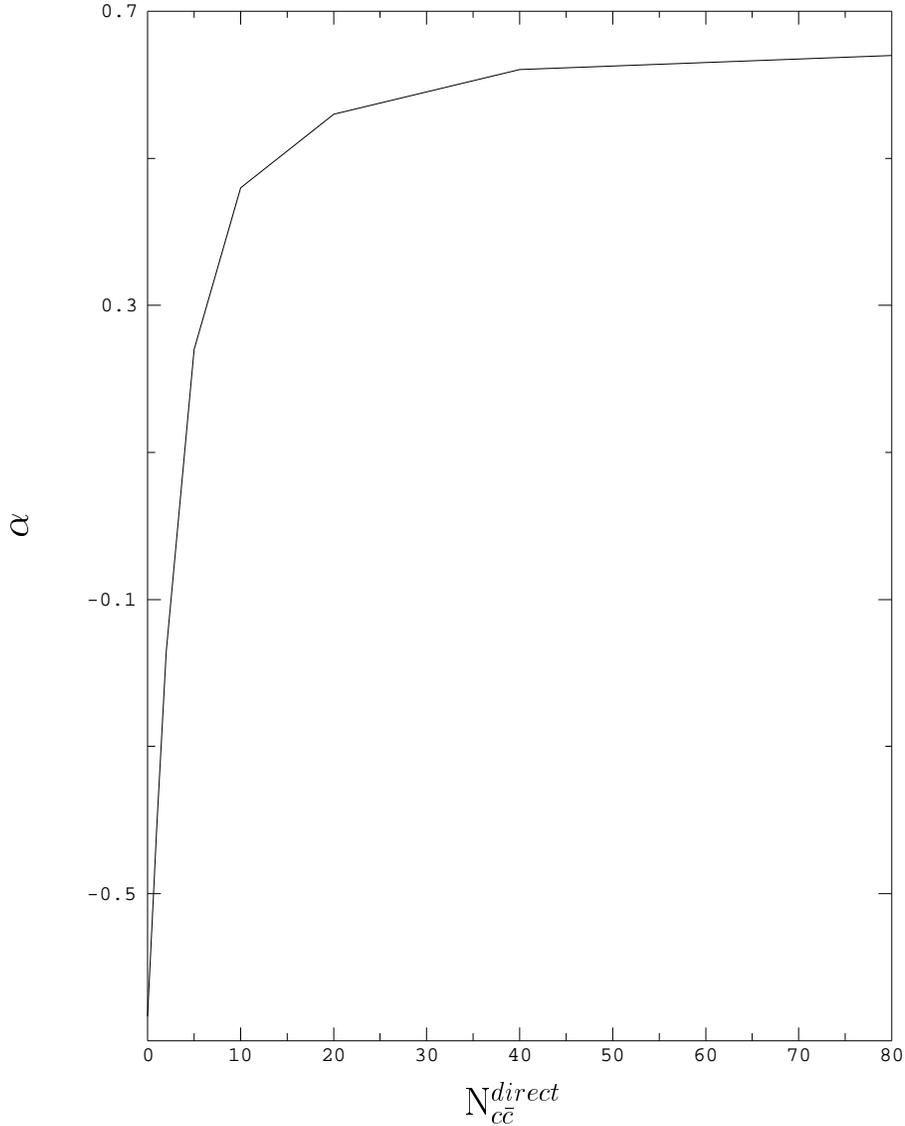}
\end{center}

\vspace{-5cm}

\caption{Dependence of $\alpha$ on the magnitude of the direct open
charm yield. For details see text. 
}
\label{fig:alphas}
\end{figure}

With the above assumed $N_{part}$ dependences one can then solve Eq.~3
to determine the $N_{part}$ dependence of $g_c$ with the result that

\beq
{\frac{N_{J/\psi}}{N_{part}} \propto N_{part}^{\alpha}},
\eeq

where $\alpha = 2/3$ in the grand-canonical limit while $\alpha =
-2/3$ in the canonical limit. From the smooth behavior of the Bessel
functions it is clear that the coefficient $\alpha$ develops
smoothly\footnote{Numerical solution indicates a slight $N_{part}$
dependence of $\alpha$ for intermediate values of N$_{c\bar
c}^{direct}$.} and continuously from -2/3 to +2/3 as N$_{c\bar
c}^{direct}$ increases from values $\ll 1$ to values  $\gg 1$.  The
numerically obtained dependence of $\alpha$ on N$_{c\bar c}^{direct}$
for fixed N$_{oc}^{therm}$ is shown in Fig.~\ref{fig:alphas}. We
conclude from this discussion 
that, in the range of validity of the statistical hadronization  
model, the centrality dependence of the yield will be
directly related to overall magnitude of the open charm cross section
relative to the thermal open charm yield.

Application of this model to the data, displayed in
Fig.~\ref{fig:psipp}, now proceeds in the following way. We start the
calculation at N$_{part} = 350$ to avoid that part of the data which
is determined by fluctuations \cite{blaizot} not contained within this
model. Furthermore, predictions of the model should only be trusted
from about N$_{part}> 150$ on, where also the $\psi'/(J/\psi)$ ratio is
close to the 
thermal value for Pb+Pb data (see Fig. 2 of ref. \cite{jpsi-stat}). In
this context it should be mentioned that the thermal character of the
$\psi'/(J/\psi)$ ratio was first noticed by \cite{shuryak}. In our
approach, ratios for all higher charmonia states including the $\chi_c$ should
approach the thermal value from N$_{part} > 150$ on, implying that for
those N$_{part}$ values feeding to J/$\psi$ should be small. In this
picture, there should thus not be different ``thresholds'' for the
disappearance of different charmonia. 

We also take the value of direct open charm production from
\cite{jpsi-stat,misko}, appropriately scaled to N$_{part}$ = 350, i.e.
$N_{c\bar c}^{direct} = 0.144$. With N$_{oc}^{therm}=0.55$ for N$_{part}
= 350$ we get, solving eq. (3), a value of g$_c$ = 1.43. Using eq. (2)
and N$_{J/\psi}^{therm}=1.84 \cdot 10^{-4}$ we obtain immediately
that $ \frac{N_{J/\psi}}{N_{part}}=1.08 \cdot 10^{-6}$ for
N$_{part}=350$ and $\alpha=-0.61$. This result is shown as the solid
line in Fig.~\ref{fig:psipp}. Considering the uncertainties in the
normalization of the J/$\psi$ and open charm yields the agreement is
surprizing. We illustrate the importance of accurately knowing the
open charm yield by the dashed line in Fig.~\ref{fig:psipp}. Here the
open charm cross section was increased by 50~\% (well within the
uncertainty of the pp data), leading to g$_c = 1.78$ and
$\alpha=-0.60$. Only a direct measurement of open charm production in
Pb+Pb collisions can remove this uncertainty of the present analysis.

Since the present analysis involves absolute densities we would like
to point out that, in the model of \cite{therm3}, an excluded volume
correction is applied. For the parameters chosen in \cite{therm3} this
leads to a density reduction by a factor of $\beta \approx 0.65$. This
should be taken into account when trying to deduce limits on open
charm production from a statistical
analysis\cite{gorenstein2}. Alternatively, one can determine the
effective volume  by requiring, as is done here for the SPS data
analysis, that the total charged particle multiplicity is correctly
reproduced by the thermal model calculations.

\section{Summary and Outlook}
\noindent Hadron production results from central nucleus-nucleus
collisions at ultra-relativistic energies can be quantitatively
understood by assuming that the fireball formed in the collision
freezes out chemically very near to the phase boundary between
quark-gluon plasma and hadron gas. 
Mesons containing heavy (charm) quarks are not thermally produced but
their yield can be quantitatively explained, at least for central
collisions, in the statistical hadronization model where all charm
quarks are directly produced  but all hadrons containing charm quarks
are formed according to thermal phase space. For small values of
$N_{c\bar c}^{direct}$ it is, as was pointed out by \cite{redlich} and
recently used by \cite{gorenstein}, important to use the canonical
approach for calculating the thermal phase space.

Predicting absolute yields for the rapidity density of J/$\psi$ mesons
at collider energies requires knowledge of the yield of open charm
pairs as well as of the total charged particle multiplicity (to fix
the volume needed for the calculation of thermal yields). The
following estimates are for central (N$_{part} = 350$) Au+Au (RHIC) or
Pb+Pb (LHC) collisions. They should be considered schematic but
illustrative of the qualitatively new features expected within the
framework presented here.  For full RHIC energy $N_{c\bar c}^{direct}$
is expected to be of order 1.5 per unit rapidity.  For a charged
particle rapidity density of dN$_{ch}/d\eta = 1000$ we get, using
eqs. (2) and (3), that $g_c=9.0$, $\alpha=-0.25$, resulting in a
J$/\psi$ yield of 0.011 per unit rapidity, close to the expected
unsuppressed value.  For LHC energies, where $N_{c\bar c}^{direct}$
approaches 50 per unit of rapidity, the grand-canonical picture should
be appropriate, i.e. $\alpha$ will be close to 2/3. The actual
computation for dN$_{ch}/d\eta = 6300$ gives $\alpha = 0.62$ and
$g_c=41$, implying a J$/\psi$ yield of about 1.4 per unit rapidity,
about a factor of 20 above the direct yield! In the grand-canonical
limit the expected J/$\psi$ yields is proportional to $(N_{c\bar
c}^{direct})^2/N_{ch}$ and can be trivially scaled to other values for
charm and charged particle production. As already discussed in
\cite{jpsi-stat}, the statistical hadronization picture should, under
these circumstances, lead to strongly enhanced quarkonia production
even compared to values for direct hard scattering production. A
similar result has been obtained recently \cite{thews} using a model
based on kinetic equations.

\section{Acknowledgements}
We  would like to acknowledge important and
stimulating discussions with K. Redlich.

\end{document}